\documentclass[prb, aps, aps,amsmath,amssymb,amsfonts, superscriptaddress,twocolumn, footinbib,longbibliography]{revtex4-2}
\usepackage[german,american,english]{babel}
\usepackage{graphicx}
\usepackage{graphics}
\usepackage{dcolumn}
\usepackage{multirow}
\usepackage{bm}
\usepackage{latexsym}
\usepackage{amssymb}
\usepackage{amsmath}
\usepackage{amsfonts}
\usepackage{layout}
\usepackage{verbatim}
\usepackage{epsfig}
\usepackage{graphicx}
\usepackage{amsbsy}
\usepackage{xcolor}
\usepackage[caption=false]{subfig}
\newcommand{\bea}{\begin{eqnarray*}}
	\newcommand{\eea}{\end{eqnarray*}}
\newcommand{\bne}{\begin{equation*}}
\newcommand{\ede}{\end{equation*}}

\newcommand{\bnen}{\begin{equation}}
\newcommand{\eden}{\end{equation}}
\newcommand{\bean}{\begin{eqnarray}}
\newcommand{\eean}{\end{eqnarray}}
\newcommand{\bsen}{\begin{subequations}}
	\newcommand{\esen}{\end{subequations}}
\newcommand{\ba}{\arraycolsep 0.3ex \begin{array}{rl}}
\newcommand{\ea}{\end{array}}
\newcommand{\bna}{\begin{array}}
	\newcommand{\eda}{\end{array}}
\newcommand{\bnm}{\begin{enumerate}}
	\newcommand{\edm}{\end{enumerate}}

\newcommand{\rom}[1]{\uppercase\expandafter{\romannumeral #1\relax}}

\newcommand {\ket} [1] {| #1 \rangle}

\def\pz{{\partial}}

\def\Hj{{\hat \jmath}}

\def\Hv{{\hat v}}

\def\HBr{\hat{\bm r}}

\def\HL{{\hat L}}

\def\Bk{{\bm k}}

\def\BE{{\bm E}}
\def\HH{{\hat H}}
\def\CR{{\mathcal R}}
\def\BCR{{\bm\CR}}
\def\Hrho{{\hat\rho}}

\def\eps{\epsilon}
\def\ve{{\varepsilon}}

\def\frac#1#2{{\textstyle{#1 \over #2}}}

\def\Der#1#2{{D #1\over D #2}}

\def\nd{^{\vphantom{\dagger}}}
\def\ns{^{\vphantom{*}}}

\def\half{\frac{1}{2}}

\def\hh{\hskip 0.1em}

\def\ket#1{{\big| \hh #1\hh \big\rangle}}

\begin{document}
	
\title{Orbital Hall effect in spin-3/2 hole-doped semiconductors and its implications for orbitronics}

\author{James H. Cullen}
\thanks{Author to whom correspondence should be addressed: james.cullen@unsw.edu.au}
\affiliation{School of Physics, The University of New South Wales, Sydney 2052, Australia}

\author{Zhanning Wang}
\affiliation{School of Physics, The University of New South Wales, Sydney 2052, Australia}

\author{Dimitrie Culcer}
\affiliation{School of Physics, The University of New South Wales, Sydney 2052, Australia}

\begin{abstract}
State-of-the-art magnetic devices rely on faster, more efficient memory elements. A major recent advance is the discovery of orbital torques, which use the orbital angular momentum of Bloch electrons to switch the magnetisation of an adjacent ferromagnet, motivating the search for orbitronic materials with strong orbital responses, exemplified by the orbital Hall effect (OHE). Here we propose $p$-type semiconductors, with a focus on Ge, as orbitronic platforms. We demonstrate that bulk holes in five common semiconductors exhibit a large orbital Hall conductivity of order $10^3 (\hbar/e)\Omega^{-1}$cm$^{-1}$, exceeding the spin-Hall effect by 2-3 orders of magnitude. The calculation is performed within the framework of the modern theory of orbital magnetisation, while incorporating recently-discovered quantum corrections to the OHE. Moreover, we argue that bulk $p$-type Ge and Si serve as ideal testbeds for the orbital torque resulting from a charge current, since the spin- and orbital-Edelstein effects are forbidden by symmetry. Our results provide a blueprint for producing strong orbital torques in magnetic devices with $p$-type semiconductors, guiding experimental work in this direction.
\end{abstract} 

\date{\today}
\maketitle


\section{Introduction}

Orbital dynamics in condensed matter systems have come under renewed scrutiny in recent years with an intense focus on out-of-equilibrium phenomena broadly encompassed by the field of orbitronics \cite{Orbitronics-in-action, Rhonald-Rev, OC-Rev-EL-2021-Yuriy, wang2024orbitronics}. The study of orbitronic phenomena involving the electrical generation and transport of Bloch electrons' orbital angular momentum (OAM) has witnessed significant experimental progress \cite{OT-FM-PRB-2021-YoshiChika, OT-OEE-NatComm-2018-Haibo,OT-PRR-2020-Hyun-Woo,OT-NatComm-2021-Kyung-Jin, Exp-OT-PRR-2020,Exp-OT-CommP-2021-Byong-Guk,OOS-Cvert-2020-PRL-Mathias, OHE-Hetero-PRR-2022-Pietro, OHE-OT-large, L-S-OT-2023,OHE-Binghai, tokura2019magnetic, go2023long, Niels_Orbitalsplitter}. This is chiefly motivated by the notion of an orbital torque, that is, a torque exerted on an adjacent magnetisation by a non-equilibrium OAM density, and is regarded as an orbital analogue of the various spin torque mechanisms that have received considerable attention in magnetic systems \cite{sakai2014, yasuda2017current, tatara2007spin, kohno2006microscopic, belashchenko2019first, CI-SOT-RMP-2019-Manchon, nikolic2020first, gambardella2011current, PhysRevB.92.014402, PhysRevB.75.214420, brataas2012current, PhysRevB.88.085423, PhysRevB.91.214401}. One of the principal mechanisms responsible for the orbital torque is the orbital Hall effect (OHE), which represents a flow of OAM in response to an electric field \cite{Orbitronics-PRL-2005-Shoucheng,ISHE-IOHE-PRB-2008-Inoue,OHE-PRL-2009-Inoue}. The OHE has been actively studied recently spurred by its potential application in magnetic memory devices \cite{Exp-OHE-Ti-Nat-2023-Hyun-Woo,Hong-OHE-PRL,OHE-PRB-2022-Manchon,OHE-metal-PRM-2022-Oppeneer,OHE-BiTMD-PRL-2021-Tatiana,OH-phase-TMD-PRB-2020-Tatiana,RS-OHE-disorder,OHE-Hetero-PRR-2022-Pietro,Inverse-OHE-weak-SOC, PhysRevLett.131.156702, PhysRevB.106.184406, PhysRevLett.131.156703,10.1063/5.0106988, bony2025quantitative, el2023observation, PhysRevB.108.245105,PhysRevB.107.094106,canonico2020two,cullen2025giant,PhysRevLett.132.106301-Giovanni,PhysRevB.111.075432,sun2024theory,IOHE-PRB-2021-Giovanni,BiTMD-OHE-PRB-2022-Giovanni&Tatiana,veneri2024extrinsic,ISOHE-PRL-2018-Hyun-Woo,IOHE-Metal-PRB-2018-Hyun-Woo,lee2025universal,santos2024negative}.

In light of the intense activity on orbital dynamics the overarching technological question concerns which mechanisms and which materials help us maximise orbital torques on ferromagnetic memory elements. In this connection we recently showed that the bulk states of topological insulators give rise to a large OHE in an electric field, building on earlier work demonstrating that orbital and spin effects in topological insulator surface states are of a similar order of magnitude \cite{PhysRevLett.108.046805}. In general, orbital and spin effects cannot be distinguished experimentally, and at the moment the only indication of their relative magnitudes comes from theoretical calculation. Moreover, whereas the equilibrium OAM in a clean system is well understood \cite{Theroy-OM-PRL-2007-Qian, Resta-PhysRevLett.95.137205, Vanderbilt_2018, Resta-PhysRevResearch.2.023139, Resta-PhysRevB.74.024408, thonhauser2011review}, most interest in the OAM at present is motivated by its out-of-equilibrium properties \cite{ Titov_EdgeOM, Exp-OEE-PRL-2022-Jinbo, lee2024orbital, OM-metal-PRB-2021-Xiaocong,pezo2024theory}, where fundamental issues need to be resolved. As an example, it was shown recently that quantum corrections to the OHE can overwhelm the conventional terms, while disorder can dominate the effect in certain systems\cite{veneri2024extrinsic,Hong-OHE-PRL}. An important fundamental question concerns inversion symmetric systems, on which there has been a significant body of work \cite{IOHE-XIV-PRB-2021-Hyun-Woo}, motivated in part by recent experiments \cite{Exp-OHE-Ti-Nat-2023-Hyun-Woo,Inverse-OHE-weak-SOC}. 

The modern theory of orbital magnetisation shows that orbital magnetisation of charge carriers in solids can be constructed via the Berry connection of Bloch wavefunctions. This theory was derived using both semiclassical and Wannier approaches \cite{Theroy-OM-PRL-2007-Qian, Resta-PhysRevLett.95.137205, Vanderbilt_2018, Resta-PhysRevResearch.2.023139, Resta-PhysRevB.74.024408, thonhauser2011review}, and has been used to describe the orbital magnetisation both in and out of equilibrium \cite{thonhauser2011review, OM-metal-PRB-2021-Xiaocong, malashevich2010theory}. Theoretical research on OAM dynamics in inversion symmetric systems has employed approaches that rely either on symmetry or on the atomic OAM operator, with virtually no work, to our knowledge, within the framework of the modern theory of orbital magnetisation. This is perhaps because the modern theory makes an explicit connection between the equilibrium OAM and the Berry curvature, which has been studied overwhelmingly in systems that break inversion symmetry. To summarise, an active search is underway for efficient orbitronic materials, while inversion symmetric systems have received little attention in the context of the modern theory of orbital magnetisation.


In light of these outstanding questions, in this work we propose bulk $p$-type semiconductors as promising platforms for orbitronic applications. We demonstrate that holes in Si, Ge, GaAs, InAs and InSb exhibit a large OHE with orbital Hall conductivities of order $10^3 (\hbar/e)\Omega^{-1}$cm$^{-1}$, a similar order of magnitude to the spin Hall effect in Pt and the orbital Hall effect recently observed in light metals \cite{ma2024spin, IOHE-Metal-PRB-2018-Hyun-Woo}. Additionally, we find that the OHE in spin-3/2 hole systems exceeds the spin-Hall effect (SHE) by 2-3 orders of magnitude. Of these semiconductors, we believe hole-doped Ge to have particular promise as an orbitronic material. Aside from proximity to Si microfabrication, ensuring high sample quality comparable to Si, hole-doped Ge has the additional advantages of ultrahigh mobilities \cite{Borsoi2025nature, stehouwer2023germanium} and considerably stronger spin-orbit coupling than Si, making it a material of choice for electrically-operated semiconductor quantum computing.\cite{scappucci2021germanium, fang2023recent} Additionally, a recent experiment has demonstrated a large inverse orbital Hall effect in Ge using YIG/W/Ge and YIG/Pt/Ge heterostructures \cite{santos2024negative}. The results indicate that the magnitude of the OHE effect in Ge is of a similar order of magnitude to the SHE in Pt, which is consistent with our results, and suggests similar magnitudes may be achieved in the other semiconductors investigated here. 

Our findings have several implications for orbitronics. Firstly, they identify a set of common $p$-type semiconductors as promising orbitronic materials. In this context, Ge and Si holes are optimal systems to test the size of the orbital torque arising from the OHE experimentally, because the spin- and orbital-Edelstein effects are prohibited by symmetry in the bulk of these materials. This implies that only the OHE and SHE are present, and our calculation, treating OHE and SHE on the same footing, shows that OHE $\gg$ SHE. Secondly, our work shows that the OHE is present in inversion-symmetric systems even within the framework of the modern theory of orbital magnetisation. The modern theory is of course general, but has overwhelmingly been applied to systems breaking inversion symmetry, which often ensure a sizable Berry curvature. Nevertheless, in the Luttinger model, even in the spherical approximation, the Berry curvature and OAM are both finite at a given wave vector, even though their integrals over occupied states naturally vanish in the absence of time-reversal breaking mechanisms. The finite Berry curvature and OAM are responsible for the large OHE that we identify. Our results are a similar order of magnitude to recent computational studies for Ge and Si, \cite{IOHE-XIV-PRB-2021-Hyun-Woo} although a direct comparison is somewhat difficult given the rather different methodology used in earlier approaches, which have focused on the atomic centred approximation. Unifying these different perspectives on a technical level will be an important undertaking for future studies. Finally, from a technical perspective, our results highlight once more the necessity of incorporating quantum corrections in the evaluation of the OHE, showing that once more the correction is larger than the conventional contribution.

A short summary of this work is as follows: We calculate the orbital and spin Hall effects in the 4 \& 6 band Luttinger-Kohn-Bir-Pikus Hamiltonian. We calculate both the conventional contribution and quantum corrections to the orbital current \cite{liu2025quantumOHE}, showing that the quantum corrections dominate. We find that the orbital Hall conductivity in spin-3/2 hole systems exceeds the spin Hall conductivity by 2-3 orders of magnitude, and we propose p-type Si, Ge, GaAs, InAs and InSb as candidates for building orbitronic devices.


\begin{figure}[tbp]
\begin{center}
\textbf{Ge dispersion}
\includegraphics[trim=0cm 0cm 0cm 0cm, clip, width=\columnwidth]{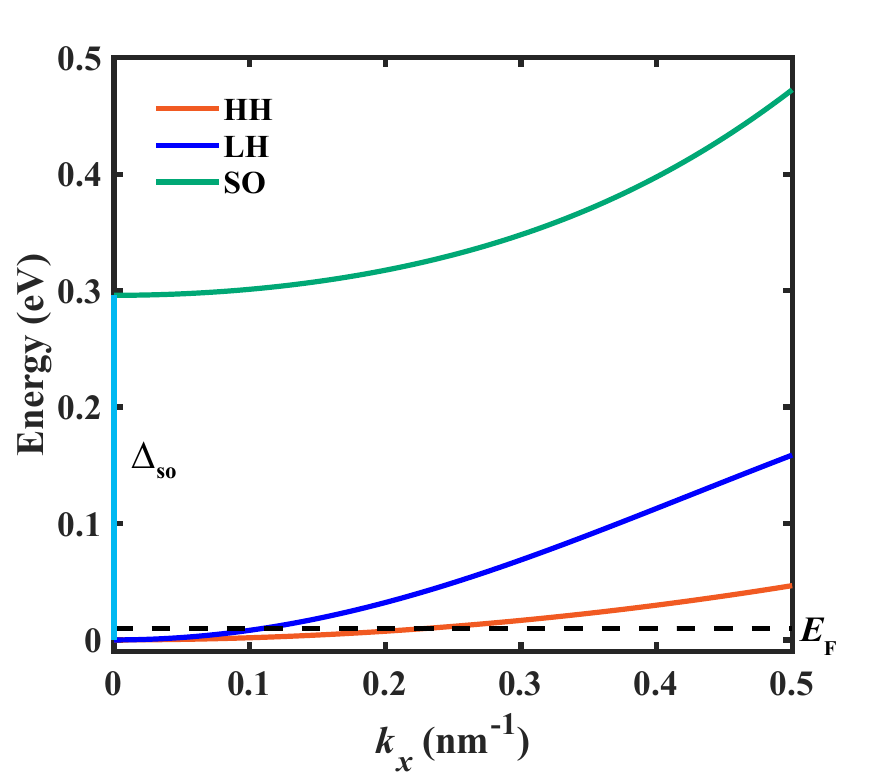}
\caption{\label{Fig:Dispersion}{ 
Dispersion for holes in Ge, showing the heavy (HH), light (LH) and split-off (SO) hole bands. In this figure the bands have been inverted so the energy is positive. Here we have chosen a Fermi energy of $10$ meV, which is the largest Fermi energy we consider in this work.}}
\end{center}
\end{figure}

\section{Results and Discussion}

\subsection{Hamiltonian}

The valence band of diamond and zinc-blende semiconductors can be described by the Luttinger-Kohn-Bir-Pikus Hamiltonian given in Refs.~\onlinecite{luttinger1955motion,Roland}:
\begin{equation}\label{Eq:6x6model}
\begin{aligned}
    &H_{0}=\\
    &\begin{pmatrix}
        P+Q & S & R & 0 & -\frac{1}{\sqrt{2}}S & -\sqrt{2} R\\
        S^* & P-Q & 0 & R & \sqrt{2}Q & \sqrt{\frac{3}{2}}S\\
        R^* & 0 & P-Q & -S & \sqrt{\frac{3}{2}}S^* & -\sqrt{2}Q\\
        0 & R^* & -S^* & P+Q & \sqrt{2} R^* & -\frac{1}{\sqrt{2}}S^*\\
        -\frac{1}{\sqrt{2}}S^* & \sqrt{2}Q & \sqrt{\frac{3}{2}}S & \sqrt{2} R & P-\delta_{\text{so}} & 0\\
        -\sqrt{2} R^* & \sqrt{\frac{3}{2}}S^* & -\sqrt{2}Q & -\frac{1}{\sqrt{2}}S & 0 & P-\delta_{\text{so}}
    \end{pmatrix}
\end{aligned},
\end{equation}
where $m_0$ is the bare electron mass, $P=-\hbar^2\gamma_1k^2/2m_0$, $Q=\hbar^2\gamma_2/2m_0(2k_z^2-k_{\parallel}^2)$, $S=\sqrt{3}\hbar^2\gamma_3k_zk_-/m_0$ and $R=\sqrt{3}\hbar^2/2m_0(\gamma_2(k_x^2-k_y^2)-2i\gamma_3k_xk_y)$, where $\gamma_{1,2,3}$ are the Luttinger parameters. This Hamiltonian describes 6 bands, 4 are degenerate at $\bm k =0$ these correspond to states with total angular momentum $3/2$, the other 2 bands are separated by a large gap $\delta_{\text{so}}$ due to the spin-orbit coupling of the atomic orbital $\xi \bm L\cdot\bm s$ and correspond the spin degenerate states with total angular momentum $1/2$. The Hamiltonian $H_0$ is a $\bm k \cdot \bm p$ Hamiltonian whose matrix elements, as well as the Luttinger parameters, are determined by the $\langle u|\bm p|m\rangle$ matrix elements of the basis states at the band edge ($\bm k=\bm 0$). The Hamiltonian (\ref{Eq:6x6model}) is written in the basis $\{|\frac{3}{2},\frac{3}{2}\rangle,|\frac{3}{2},\frac{1}{2}\rangle,|\frac{3}{2},-\frac{1}{2}\rangle,|\frac{3}{2},-\frac{3}{2}\rangle,|\frac{1}{2},\frac{1}{2}\rangle,|\frac{1}{2},-\frac{1}{2}\rangle\}$ where the first number indicates the total angular momentum and the second number indicates its projection onto $J_z$.\cite{Roland}

Holes with their total angular momentum equal to $3/2$ and $J_z$ projection equal to $\pm3/2$ at the band centre are known as heavy holes while those with $J_z=\pm1/2$ are known as light holes due to their different effective masses shown in Fig.~\ref{Fig:Dispersion}. This figure also shows the two bands with total angular momentum equal to $1/2$ with gap $\delta_{\text{so}}$ -- these are referred to as the split-off bands. The split-off band gap energies for each of the semiconductors considered in this work are given in Tab.~\ref{tab:split-off}. For materials with a large split-off band gap such as Ge, GaAs, InAs and InSb, we can simply use the upper $4\times4$ block from (\ref{Eq:6x6model}) to describe the system,\cite{chow&koch} given that we only consider small number densities/Fermi energies as indicated in Fig.~\ref{Fig:Dispersion}. We note that the effective spin-$3/2$ of hole systems in such $4\times4$ models leads to dynamics that either has no equivalent in electron systems \cite{culcer2006spin} or is very difficult to observe in electron systems.\cite{liu2018strong} For describing Si the $6\times 6$ model is the minimum requirement due to the smaller split-off band gap ($\sim 40$ meV) and the large relative difference between the $\gamma_2$ and $\gamma_3$ Luttinger parameters. At the same time, qualitative insight can still be gained by considering the effective spin-3/2 dynamics of the top valence bands.

Additionally, we can also use the spherical approximation, which works well for Ge, GaAs, InAs and InSb, and reduces the effective Hamiltonian to
\begin{equation}\label{Eq:4x4model}
    H_0 = -\frac{\hbar^2}{2m}\left[(\gamma_1+\frac{5}{2}\bar{\gamma})k^2-2\bar{\gamma}(\boldsymbol{k}\cdot\boldsymbol{J})^2\right]\,,
\end{equation}
where $J_i$ are the spin $3/2$ matrices and $\bar{\gamma}=(\gamma_2 + \gamma_3)/2$. The $4\times4$ spherical model above allows for the straightforward analytical calculation of the orbital current. In this work employ the $4\times4$ model with and without the spherical approximation, primarily for analytical insight, as well as the full $6\times6$ model for numerical accuracy. For the $4\times4$ model in the spherical approximation we calculate the orbital and spin Hall conductivities analytically, whereas for the other two models we calculate orbital Hall conductivity numerically. 

\subsection{Equilibrium OAM and Berry curvature}
The Berry curvature for band $m$ and wavevector $\bm k$ is defined as $\Omega^m_{i,\bm k} = \epsilon_{ijl}{\rm Im} \langle \partial_j u_{m\bm k}| \partial_l u_{m\bm k}\rangle$ and for { 3D holes in the spherical Luttinger model it takes the form}
\begin{equation}
    \Omega^m_{i,\bm k} = -\frac{k_i}{k^3}J_z^{mm}\,.
\end{equation}
The Berry curvature is closely related to the equilibrium OAM. The OAM operator is defined as the symmetrised combination $\hat{\bm L}=\frac{m}{2}(\hat{\bm r}\times\hat{\bm v} - \hat{\bm v}\times\hat{\bm r})$, where $\hat{\bm v}$ is the velocity operator and $\hat{\bm r}$ the position operator. Note that the mass appearing in the OAM operator is the bare electron mass, hence for holes we use the negative of the electron mass. The local circulation part of the equilibrium OAM for band $m$ and wavevector $\bm k$ is calculated as $L^{m}_{i,\bm k} =-\frac{m}{\hbar}\epsilon_{ijl} {\rm Im} \langle \partial_j u_{m\bm k }|\epsilon^{m}_{\bm k}-H_0| \partial_l u_{m\bm k }  \rangle $\cite{Theroy-OM-PRL-2007-Qian, Resta-PhysRevLett.95.137205, Vanderbilt_2018, Resta-PhysRevResearch.2.023139, Resta-PhysRevB.74.024408, thonhauser2011review}. The orbital angular momentum of 3D holes in band $m$ and wave vector $\bm k$, retaining the spherical approximation, is
\begin{equation}
    L^{m}_{i,\bm k} = \frac{3 \hbar \bar{\gamma} k_i}{k}(\sigma_z\otimes\mathbb{I})^{mm}\,,
\end{equation}
where $\sigma_z$ is the $z$ Pauli matrix and $\mathbb{I}$ is the $2\times2$ identity matrix. It is evident from the above that, despite the presence of inversion symmetry in the Luttinger model, the Berry curvature of each band does not vanish, and the OAM is finite for a hole in band $m$ with wavevector $\bm k$. However, the integral of both the OAM and Berry curvature over occupied states will vanish as expected, since the system has time reversal symmetry. { These results remain true if the spherical approximation is removed and the model is extended to $6 \times 6$, as we do in the remainder of this work, except the results can no longer be written in a simple and revealing analytical form.}

\subsection{Non-equilibrium formalism}

To evaluate the orbital and spin Hall effects we require the non-equilibrium correction to the density matrix in an electric field, for which we use the linear response theory following the approach of Refs.~\onlinecite{Interband-Coherence-PRB-2017-Dimi, JE-PRR-Rhonald-2022}. The single-particle density operator obeys the quantum Liouville equation, $\pz\Hrho/\pz t + (i/\hbar)[\HH,\Hrho]=0$, where $\HH=\HH_0+\HH_E$. Here $\HH_0$ is the band Hamiltonian and $\HH_E=e\BE\cdot\HBr$ is the potential due to the external electrical field. We work in the Hilbert space spanned by Bloch wave-functions $\ket{\Psi_{m{\bm k}} } = e^{i{\bm k}\cdot{\bm r}} \ket{u_{m{\bm k}}}$. In the crystal momentum representation the equilibrium density matrix has the diagonal form $\rho_{0\Bk}^{mn} = f_{m\bm k}\, \delta_{mn}$, where $f_{m\bm k} \equiv f(\ve_{m{\bm k}})$ is the Fermi-Dirac distribution for band $m$. In an electric field the density matrix can be written as $\hat{\rho} = \rho_0+\rho_E$, and, in linear response, it has been shown that in the absence of disorder\cite{Interband-Coherence-PRB-2017-Dimi} 
\begin{equation}\label{rhoE}
\rho^{mn}_{E{\bm k}} = {f(\ve_{m{\bm k}})-f(\ve_{n{\bm k}})\over
\ve_{m{\bm k}} -\ve_{n{\bm k}} }\> e{\bm E}\cdot \BCR^{mn}_\Bk\quad,
\end{equation}
where $\boldsymbol{\mathcal{R}}_{\bm k}^{mn}=\langle u_{n\bm k}|i\partial u_{m\bm k}/\partial \bm k\rangle$ is the Berry connection. { In this work we do not consider disorder-induced extrinsic terms and instead focus only on the intrinsic contributions.}

Now, our evaluation of the orbital current follows the calculation in Ref.~\onlinecite{liu2025quantumOHE}. The orbital current operator is defined as $\Hj^\alpha_\delta = \half\big\{\HL_\alpha,\Hv_\delta\big\}$, where the OAM polarization is taken to be along the $\alpha$-direction while the transport direction is denoted by $\delta$. The expectation values of $\Hj$ is then evaluated by taking the trace with the density matrix. Once $\rho^{mn}_{E{\bm k}}$ is found the expectation value of the orbital current can be written as
\begin{equation}\label{OC}
\begin{aligned}
    &\langle\Hj^\alpha_\delta\rangle =\frac{m\eps\ns_{\alpha\beta\gamma}}{4}  \sum_{m,\Bk} \big\{\CR\nd_\beta, \rho\nd_{E\Bk}\big\}^{mm}\, \big\{v_\delta, v_\gamma\big\}^{mm} +\\
    &i\frac{m\eps\ns_{\alpha\beta\gamma}}{4} \sum_{m \ne n,\Bk}{2eE_\mu \left[\Der{\Xi^0_\beta}{k_\mu}\right]^{mn} +\big\{\hbar v\nd_\beta, \rho\nd_{E\Bk}\big\}^{mn} \over \ve_n-\ve_m} \{v_\gamma, v_\delta\}^{nm}\\
    &+i\frac{m\eps\ns_{\alpha\beta\gamma}}{4} \sum_{m \ne n,\Bk}\Big[v_\gamma, \Der{v_\delta}{k_\beta} \Big]^{mn}_\Bk \rho^{nm}_{E\Bk} \,,
\end{aligned}
\end{equation}
where $m$ and $n$ are band indices, ${\bm E}$ is the external electric field, $\big[\Xi^0_\beta\big]^{mn}=\half\CR^{mn}_\beta (f_m + f_n)$, and the covariant derivative $DO/Dk_j=\partial O/\partial k_j - i[\mathcal{R}_j, O]$. This orbital current expression (\ref{OC}) was derived in Ref.~\onlinecite{liu2025quantumOHE} and shown to be gauge invariant. The expression in (\ref{OC}) contains the quantum correction to the orbital current $\Delta j$ that arises due to the inclusion of all matrix elements, intra-band and inter-band, of the position and velocity operators. The conventional part of the orbital current is contained in the first term of (\ref{OC}), but only contains the off-diagonal components of the velocity operators, while the quantum correction comprises all the other terms in (\ref{OC}). { We note that the conventional calculation of the orbital current only includes diagonal elements of the OAM operator in the trace $\langle J_\delta^\alpha\rangle=\mathrm{Tr}\half\big\{\HL_\alpha,\Hv_\delta\big\} \rho$, whereas the quantum correction accounts for the remainder of the matrix elements appearing in this trace. The conventional approach does not have a physical justification for ignoring these terms. Additionally, the quantum corrections include terms corresponding to both local and itinerant circulation, so these corrections are also important in materials with their OAM localised about the atomic centre.} In the limit where $\bar{\gamma}\rightarrow0$ all states become degenerate and the intrisinc part of the nonequilibrium density matrix $\rho_{E\Bk}$ becomes zero, as such in this limit the orbital current trivially vanishes.

The quantum correction $\Delta j$ can be split into three contributions $\Delta j_{1,2,3}$ \cite{liu2025quantumOHE}. The first contribution $\Delta j_1$ can be related to the generation of an inter-band polarization by an applied electric field, displacing electrons away from their equilibrium center of mass. This dipole rotates, generating an OAM, and the OAM is then convected generating an orbital current. This mechanism can also be used to describe the conventional contribution. The first quantum correction $\Delta j_1$ and the second $\Delta j_2$ are the two most dominant contributions to the orbital current in holes. $\Delta j_2$ arises due to the interband matrix elements of the OAM operator, these elements represent the components of the OAM that fluctuate with time. While these matrix elements do not contribute to the expectation value of the OAM in equilibrium, they do contribute to the orbital current. The last contribution to quantum correction $\Delta j_3$ arises due to the non-commutativity of the position and velocity operators. For both the $4\times 4$ and $6\times6$ Luttinger models, $\Delta j_3$ has opposite sign to all other contributions to the orbital current.

Expressions for the proper spin Hall current have been derived in Refs.~\onlinecite{Hong-PSHE-PRB,ma2024spin,Cong-CC-PRB} using both semiclassical and fully quantum mechanical formalisms. Here we follow the quantum mechanical formulation based on Bloch wavefunctions from Refs.~\onlinecite{Hong-PSHE-PRB,ma2024spin}. The general analytical expression for the intrinsic proper spin Hall conductivity in systems with arbitrary degeneracies is
\begin{equation}\label{CS-main}
    \sigma^l_{ij,\text{SHE}} = -\frac{2 e}{\hbar}\sum_{{\bm k}} \sum_{mnn^\prime} f(\ve_{m{\bm k}}) \text{Im}\left[\Tilde{\mathcal R}^{mn}_{i,\bm k}\check{s}^{nn^\prime}_{l,\bm k}\Tilde{\mathcal R}^{n^\prime m}_{j,\bm k}\right],
\end{equation}
where ${\bm s}$ is the spin operator. The check over the spin term indicates the inclusion of only band diagonal matrix elements and elements between degenerate states, and the tilde over the Berry connection indicates the inclusion of only matrix elements between non-degenerate states. { The distinction between the conventional and proper spin currents in the spin Hall effect is important for materials in which the spin is not conserved. A similar distinction will likely need to be made with respect to the orbital current; however, the most appropriate way to do this remains an open question in the field. It was recently shown in Ref.~\onlinecite{Rhonald-Conservation-OMM} that the OAM is conserved when there is no orbital magneto-electric effect in the bulk, and as such the definition used for the orbital current in this work is applicable to the models used.}

Finally, whereas it is not the aim of the present paper to revisit the substantial debate surrounding the spin-Hall effect, we note in passing that a separate expression for the spin-Hall conductivity has been derived in the literature. \cite{Cong-CC-PRB, valet2025quantum}
This expression is qualitatively very similar to ours and yields results of the same order of magnitude, yet differs in a number of details, which will be discussed in a future publication. Nevertheless, for the purposes of comparing OHE and SHE dynamics these distinctions are immaterial, and we expect our observations to hold regardless of the explicit form used to determine the spin current.

\begin{figure}[tbp]
\begin{center}
\textbf{Ge OHE in spherical approximation}
\hspace{3mm}
\includegraphics[trim=0cm 0cm 0cm 0cm, clip, width=0.85\columnwidth]{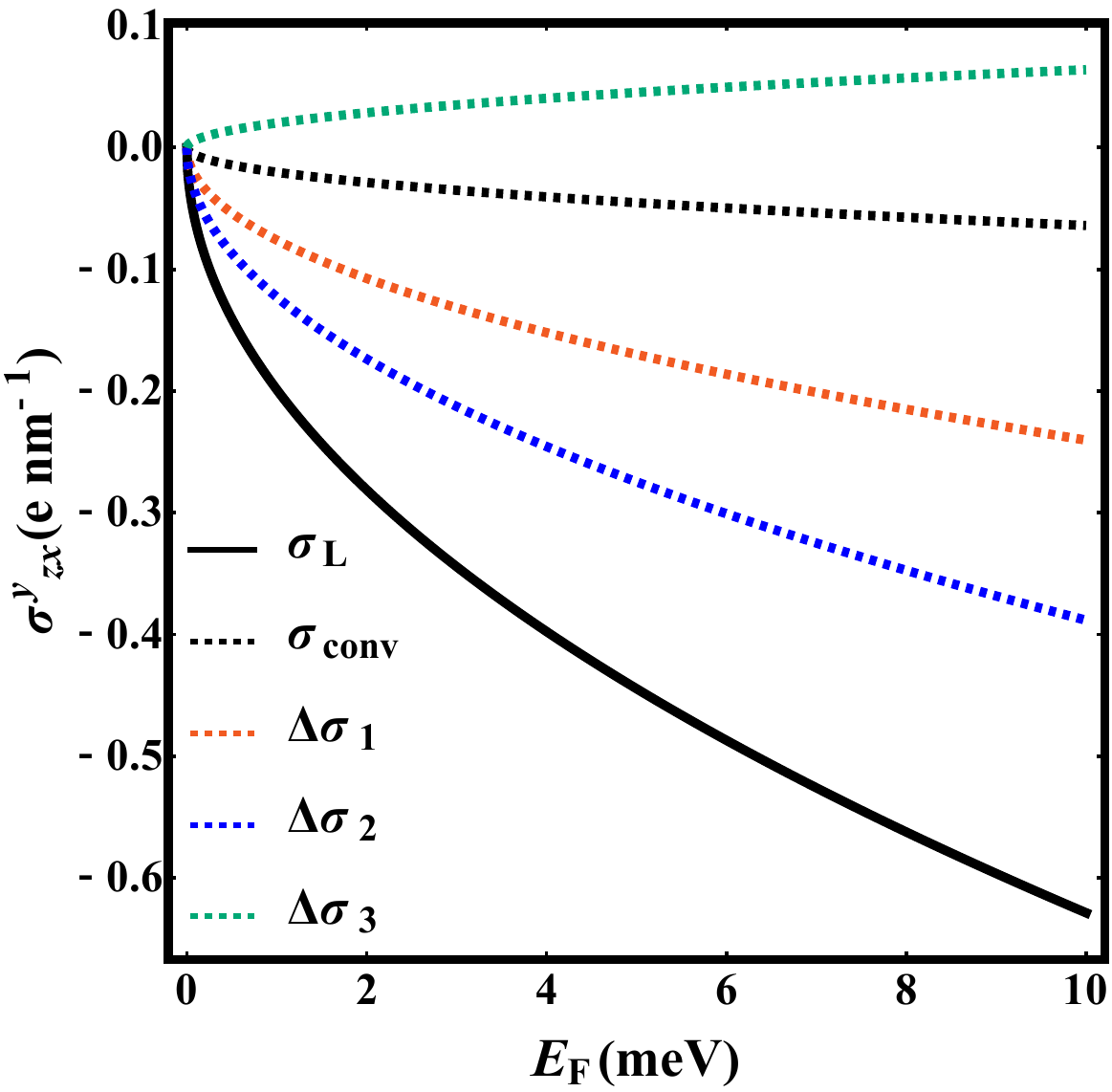}
\caption{\label{Fig:OHE_Ge}
The orbital Hall conductivity vs the Fermi energy in Ge in the spherical approximation. We have also plotted the conventional part of the orbital Hall conductivity $\sigma_{\text{conv}}$ along with the quantum corrections $\Delta\sigma$. Here we use the Luttinger parameters $\gamma_1=13.38$ and $\bar{\gamma}=4.97$.}
\end{center}
\end{figure}

\subsection{OHE calculations}
Here we present our results for the OHE, and include results for the SHE to enable a complete characterisation. We first derived analytical expressions for the orbital and spin conductivities based on the spherical $4\times 4$ model: derivations of these expressions can be found in SUPPLEMENTARY NOTES \rom{1}--\rom{3}. Additionally, we present numerical calculations of the orbital Hall conductivity based on both the $4\times 4$ model and the $6\times 6$ model without the spherical approximation.

The orbital Hall conductivity based on the analytical calculation is plotted in Fig. \ref{Fig:OHE_Ge}, the figure shows all the orbital Hall conductivity components for Ge. We find the the orbital Hall conductivity to be of the same order of magnitude to the spin Hall effect in Pt and the orbital Hall effect in light metals \cite{ma2024spin,IOHE-Metal-PRB-2018-Hyun-Woo}. Moreover, as shown in Fig.~\ref{Fig:OHE_Ge}, the quantum corrections $\Delta\sigma_{1,2}$ are the dominant contributions to the orbital Hall conductivity. The dominance of the quantum correction is consistent with our previous results in Refs.~\onlinecite{liu2025quantumOHE, cullen2025giant}, and highlights the importance of including these corrections when calculating the orbital current.

\begin{figure}[tbp]
\begin{center}
\textbf{SHE in p-type semiconductors}
\hspace{3mm}
\includegraphics[trim=0cm 0cm 0cm 0cm, clip, width=0.85\columnwidth]{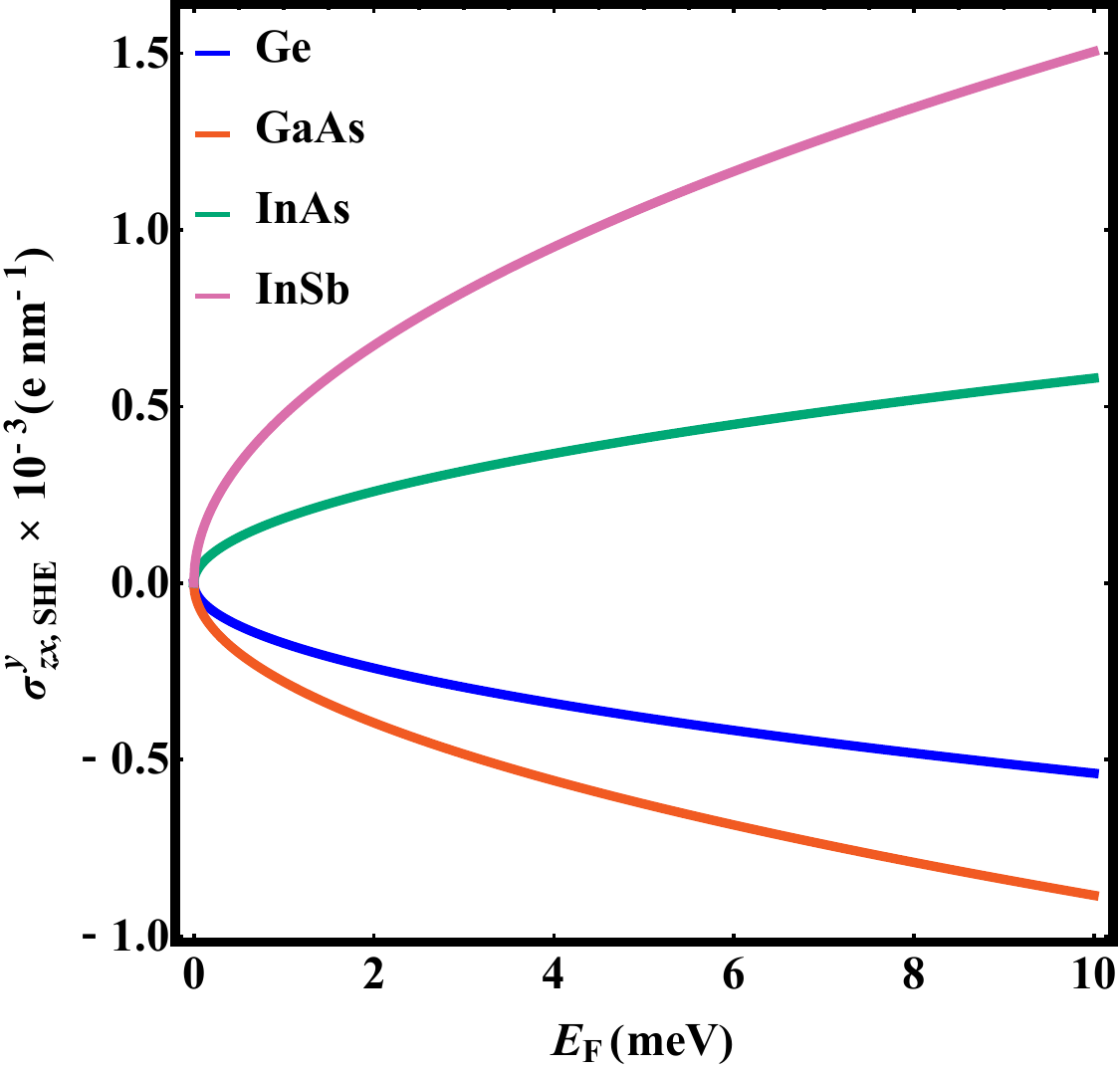}
\caption{\label{Fig:SHE}
The proper spin Hall conductivity vs the Fermi energy in for four semiconductors in the spherical approximation. Here we use the Luttinger parameters from Ref.~\onlinecite{Roland}. Note the difference of three orders of magnitude in the scale of the $y$-axis as compared to the orbital Hall conductivity.}
\end{center}
\end{figure}

We have plotted the spin Hall conductivity vs the Fermi energy in Fig. \ref{Fig:SHE}, as shown in the figure we find the spin Hall conductivity to be 2-3 orders of magnitude smaller than the orbital Hall conductivity. We have excluded Si from Fig. \ref{Fig:SHE} due to the inapplicability of the spherical approximation to this material. We find the spin and orbital conductivities to have the same sign in both Ge, Si and GaAs, while in InAs and InSb their signs are opposite. The spin and orbital conductivities having the same sign in Ge and Si differs from the ab-initio calculation of Ref.~\onlinecite{IOHE-XIV-PRB-2021-Hyun-Woo}. At the moment, however, our results and the results of Ref.~\onlinecite{IOHE-XIV-PRB-2021-Hyun-Woo} cannot be compared directly, since Ref.~\onlinecite{IOHE-XIV-PRB-2021-Hyun-Woo} introduced the OAM from the perspective of atomic orbitals, which is somewhat different from the way it is determined in the modern theory. Furthermore, for completeness we indicate that
Ref.~\onlinecite{IOHE-XIV-PRB-2021-Hyun-Woo} focussed on the conventional spin current, which can have opposite sign to the proper spin current\cite{ma2024spin}. However, the conventional and proper spin currents typically yield results that are close in magnitude.

\begin{figure}[tbp]
\begin{center}
\textbf{Ge OHE in $4\times4$ and $6\times6$ models}
\includegraphics[trim=0cm 0cm 0cm 0cm, clip, width=0.9\columnwidth]{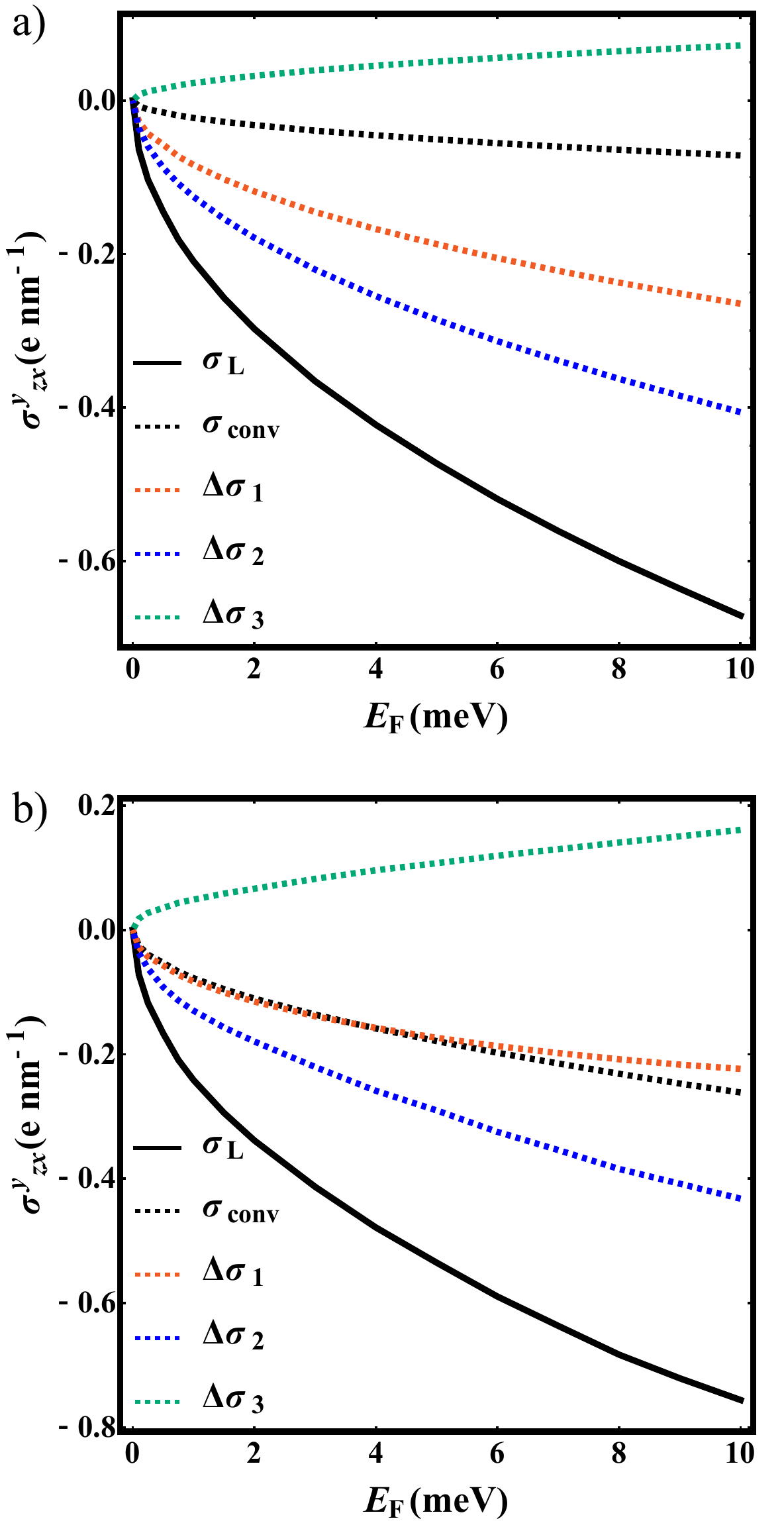}
\caption{\label{Fig:OHE_Ge_compare}
{ The orbital Hall conductivity vs the Fermi energy in Ge for a) the $4\times4$ Luttinger Hamiltonian and b) the $6\times6$ Luttinger Hamiltonian without the spherical approximation. We have plotted the conventional part of the orbital Hall conductivity $\sigma_{\text{conv}}$ along with the quantum corrections $\Delta\sigma$. Here we use $\gamma_1=13.38$, $\gamma_2=4.24$ and $\gamma_3=5.69$.}}
\end{center}
\end{figure}

We have numerically calculated the orbital Hall conductivity using the $4\times 4$ and $6\times 6$ Luttinger models without the spherical approximation to compare with our analytical $4\times 4$ calculation. A plot of the orbital Hall conductivity including all components for these models is shown in Fig.~\ref{Fig:OHE_Ge_compare} for comparison with Fig.~\ref{Fig:OHE_Ge}. As is shown, the difference between orbital Hall conductivities obtained using the spherical approximation and the regular $4\times4$ model is $8\%$ for Ge. Furthermore, we find that the inclusion of the split-off band does not significantly affect the magnitude of the orbital Hall conductivity (correction of $12\%$). As such, we do not expect the inclusion of further bands to have a significant effect on the magnitude of the orbital Hall conductivity because of the increasingly large energy separations with the hole bands. The main changes in the results between the $4\times 4$ and $6\times 6$ models are in the conventional term and the third quantum correction; however, these contributions almost exactly cancel. 

Using the full $6\times 6$ model we have plotted the orbital Hall conductivity for Ge, Si, GaAs, InAs and InSb in Fig.~\ref{Fig:OHE_allmats}. We find the orbital Hall conductivity to be of comparable magnitude in all materials studied $10^{-1}-10^0$ e nm$^{-1}$. As such, all of these semiconductors exhibit very large orbital Hall effects and have potential for use in building orbitronic devices.


\subsection{Implications for orbitronics}
\begin{figure}[tbp]
\begin{center}
\textbf{OHE in p-type semiconductors}
\hspace{3mm}
\includegraphics[trim=0cm 0cm 0cm 0cm, clip, width=0.85\columnwidth]{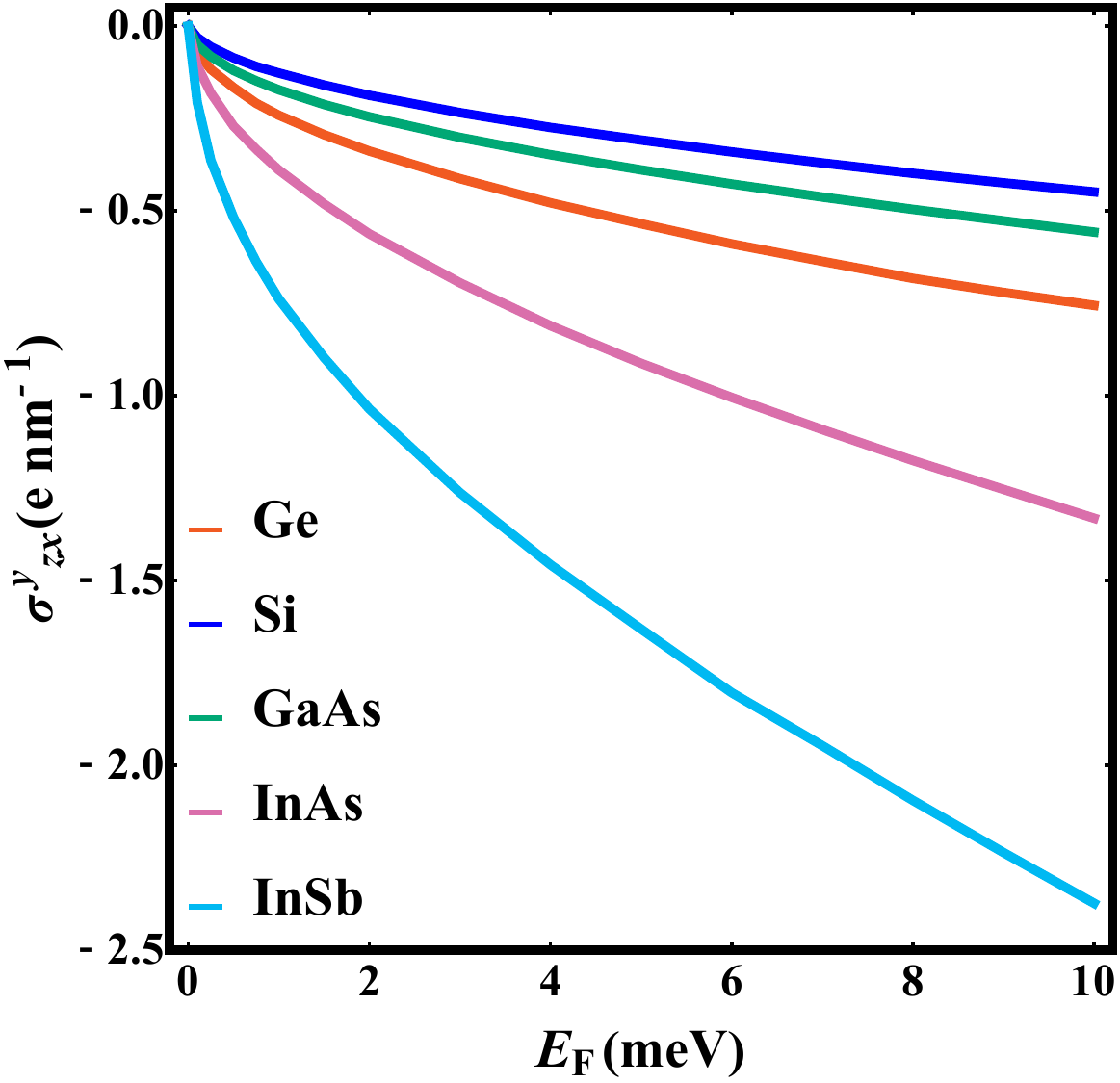}
\caption{\label{Fig:OHE_allmats}
The orbital Hall conductivity vs Fermi energy for various hole-doped semiconductors calculated using the $6\times6$ Luttinger Hamiltonian. Luttinger parameters have been taken from Ref.~\onlinecite{Roland}.}
\end{center}
\end{figure}

Our main finding is that the OHE of holes in bulk semiconductors is very large, it is 2-3 orders of magnitude larger than the SHE and is of a similar order of magnitude to the OHE calculated in weakly SOC metals in Ref.~\onlinecite{IOHE-Metal-PRB-2018-Hyun-Woo}. We have plotted the orbital Hall conductivity vs the Fermi energy for these semiconductors in Fig.~\ref{Fig:OHE_allmats}, we considered the most common semiconductors described by the Luttinger Hamiltonian: Si, Ge, GaAs, InAs and InSb. In all of these calculations we have used the $6\times 6$ Luttinger Hamiltonian without the spherical approximation and including the split-off band. Despite p-type Ge exhibiting the third largest orbital Hall conductivity of the materials studied we believe it is the best candidate for building orbitronic devices. Ge is grown to a high degree of purity resulting in ultrahigh quality samples, benefits from its proximity to Si microfabrication technologies, and exhibits substantial hole mobilities even in the bulk. The time of flight hole mobility in 3D Ge at 40K has been measured to be of the order of $10^5$ cm$^2$/(Vs) \cite{reggiani1977hole,ottaviani1973hole}, this is a similar order of magnitude to the mobility of holes in Si and a greater order of magnitude than the mobility in GaAs, InAs and InSb \cite{OttavianiSi, dalal1971GaAs, adachi2005propertiesInAs, filipchenko1976InSb, kesamanly1968InAs,zimpel1989mobilityInSb}. From this perspective, since the OHE in Ge is greater than in Si, Ge is an optimal material for orbitronic applications. We also note that, although efficient OAM injection via OHE requires a 3D structure, mobilities measured in 2D Ge structures are considerably higher than the above value, offering the prospect of further improvement\cite{Borsoi2025nature, stehouwer2023germanium}. 

\begin{figure}[tbp]
\begin{center}
\textbf{Orbital torque in a Ge device}
\includegraphics[trim=0cm 0cm 0cm 0cm, clip, width=0.95\columnwidth]{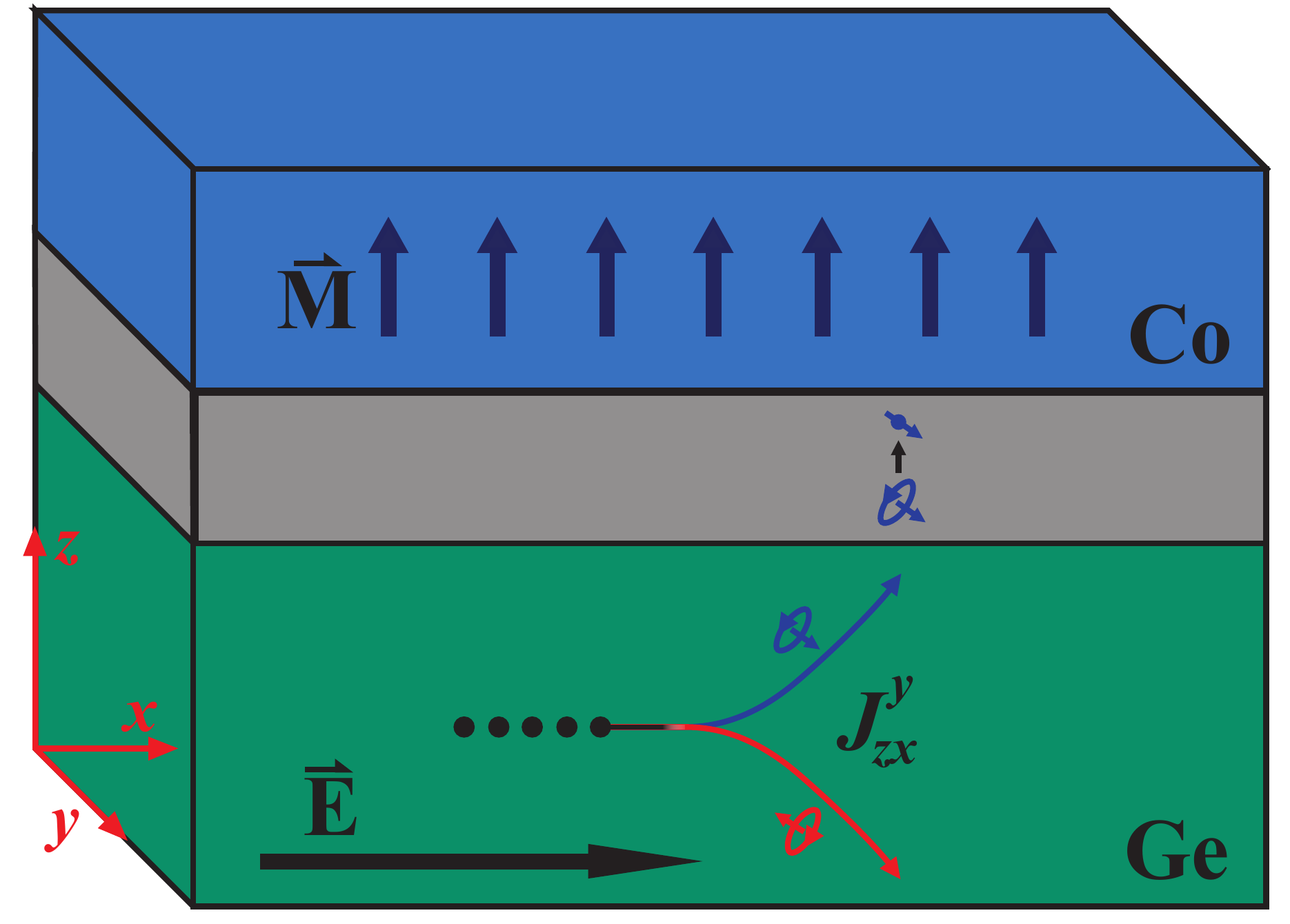}
\caption{\label{Fig:Ge_Co_OHE}
Diagram of the orbital Hall torque in a Ge/Co heterostructure. Here the applied electric field generates transverse orbital currents via the intrinsic orbital Hall effect. The orbital current will generate an orbital accumulation. The orbital angular momentum is then converted into spin which generates a torque on the magnetisation in the Co layer. A barrier layer may be used to assist with orbital-to-spin conversion.}
\end{center}
\end{figure}

Our study, taken together with the observations above, suggests the viability of Ge-metal interfaces, for example Ge/Co, where Co is a metal of choice in spintronic ferromagnetic structures \cite{bader2010spintronics, felser2007spintronics}. This interface, and its relatives, have featured in a number of studies \cite{nedelkoski2025effects, tsay2003magnetic, dhar1998atomic, tsay2007magnetic}, including a recent study linked to spin injection \cite{nedelkoski2025effects}, but we are not aware of its use for orbitronic applications. Given the difference in magnitude of the spin and orbital Hall effects we can expect that for any significant orbital-to-spin conversion $> 0.2\%$ the orbital Hall effect will dominate the torque in a Ge/ferromagnetic device. Fig.~\ref{Fig:Ge_Co_OHE} shows a provisional sketch that illustrates one possible geometry for a Ge orbital torque device. The Ge/Co device depicted in Fig.~\ref{Fig:Ge_Co_OHE} includes a hypothetical barrier layer, as these layers are often used to assist with orbital-to-spin conversion \cite{OHE-Hetero-PRR-2022-Pietro,L-S-OT-2023}, however, the barrier layer may not be necessary. In such a device the orbital Hall effect can be studied via the magnetisation dynamics in the ferromagnet \cite{mellnik2014spin}. 

\begin{table}[tbp] 
\centering
Band gaps
\hspace{3mm}
\begin{tabular}{ m{0.15\columnwidth} | m{0.15\columnwidth} m{0.15\columnwidth} m{0.15\columnwidth} m{0.15\columnwidth} m{0.15\columnwidth}}
\hline\hline
 & Ge & Si & GaAs & InAs & InSb\\
\hline
\vspace{1mm}
$\Delta_{\text{so}}$ (eV) & $0.296$ & $0.0441$ & $0.341$ & $0.380$ & $0.810$ \\
$\Delta_{\text{c}}$ (eV) & $0.8$ & $3.4$ & $1.5$ & $0.42$ & $0.24$ \\
\hline\hline
\end{tabular}
\caption{\label{tab:split-off} The split-off band energy splitting $\Delta_{\text{so}}$ and direct valence-conduction band gaps $\Delta_{\text{c}}$ for different semiconductors, values taken from Ref.~\onlinecite{Roland}.}
\end{table}

In order for an orbital current to exert a torque on a magnetisation it is thought that the OAM undergoes an orbital-to-spin conversion process, the exact mechanisms behind this phenomenon are largely unknown for all forms of the Bloch OAM. As such, methods for enhancing the orbital-to-spin conversion are still rather ambiguous. Although there is a proposed mechanism for the atomic OAM converting to spin via on-site spin-orbit coupling, this mechanism lacks rigorous theoretical justification. Additionally, the details of the transport of the Bloch OAM, both local and itinerant, and its conservation across interfaces are again open questions in the field. So, calculations of the orbital Hall conductivity are the primary indicator of whether a material can generate a large orbital torque. Our calculations for the orbital Hall effect in p-type semiconductors is the same order of magnitude as what has been theoretically predicted for Ti, however, measurements for the orbital torque in Ti/Ni bilayers have estimated the orbital Hall conductivity to be an order of magnitude smaller\cite{Exp-OHE-Ti-Nat-2023-Hyun-Woo}, this discrepancy is likely due to the combination of the orbital transparency of the interface and the orbital-to-spin conversion efficiency of the device.

\subsection{Testbed for orbital Hall torque}
Distinguishing orbital and spin effects is not possible using currently available experimental techniques, and at the moment the best one can do is to calculate these effects theoretically and determine whether one clearly dominates over the others. Similarly, distinguishing between angular momentum generated via different mechanisms is equally challenging. In addition to the spin Hall effect, a steady-state spin density can be generated in an electric field via the magneto-electric effect, also known as the Edelstein effect \cite{edelstein1990spin,aronov1989nuclear}. Likewise, the orbital magneto-electric effect refers to the intrinsic generation of a steady state orbital polarization by an electric field \cite{OEE-scalar-potential,OEE-NL-2018-Shuichi,OEE-SciR-2015-Shuichi,OEE-NatComm-2019-Peter,OAM-SciRep-2017-Yuriy,park2012orbital,OM-metal-PRB-2021-Xiaocong, johansson2024theory, osumi2021kinetic,he2020giant,Edelstein-PhysRevResearch.3.013275, PhysRevB.102.184404, canonico2025spin}. However, the spin- and orbital-Edelstein effects require gyrotropic symmetry, which is absent in zincblende materials. Given that the orbital Hall conductivity is three orders of magnitude larger than the spin Hall conductivity in Ge holes and two orders larger in Si holes, this suggests that in the bulk virtually all the angular momentum dynamics comes from the OHE. The dynamics near the interface need to be examined separately, since other mechanisms may contribute to angular momentum accumulation there \cite{valet2025quantum}, nevertheless we expect OHE to be a strong contributor there. Hence, our results indicate that 3D Ge and Si structures are suitable testbeds for the strength of the orbital Hall torque\cite{BoseOrbitltorquePRB}. Additionally, it was recently demonstrated that $n$-type Si can exhibit a sizable orbital torque\cite{matsumoto2025observation}. Due to the larger spin-orbit coupling in holes compared with electrons in Si, we expect $p$-type Si to potentially exhibit an even larger orbital torque. 


\subsection{Inversion symmetric structures and the modern theory}
The primary technological motivation of the present work is to uncover the orbitronic properties of p-type semiconductors by determining the OHE in of holes within the framework of the modern theory of orbital magnetisation, incorporating the quantum corrections, and studying the OHE and SHE on the same footing. At the moment there is no study of the OAM and OHE for the Luttinger Hamiltonian in the modern theory. A strong theoretical motivation for working within the modern theory is the need to understand (i) the OHE in inversion-symmetric spin-3/2 systems within this framework, as well as (ii) the role of spin-orbit coupling in giving rise to the OHE.

To illustrate this need we emphasise that our results are in sharp contrast to the earlier method introduced in Ref.~\onlinecite{Orbitronics-PRL-2005-Shoucheng}, where a nominally similar calculation was performed for Si holes, which, however, neglected spin-orbit coupling. The Hamiltonian of Ref.~\onlinecite{Orbitronics-PRL-2005-Shoucheng} was expressed in terms of a pseudo-OAM operator, and the transport of this pseudo-OAM was investigated in the absence of spin-orbit coupling, yielding finite results. Nevertheless, if one were to calculate the OAM for this effective Hamiltonian according to the modern theory the result will immediately seen to be zero since spin-orbit coupling is absent. Within the modern theory the OAM in the Luttinger Hamiltonian is entirely due to the spin-orbit interaction encapsulated in the Luttinger parameters $\gamma_2$ and $\gamma_3$. Hence one cannot speak of OHE, or any OAM effects, in a hole gas in the modern theory in the absence of spin-orbit coupling. This is consistent with the findings of a recent theory of angular momentum transport. \cite{valet2025quantum} It follows that the pseudo-OAM contained in the basis envelope functions of Ref.~\onlinecite{Orbitronics-PRL-2005-Shoucheng}, which can be regarded as a parent Hamiltonian of the Luttinger Hamiltonian, is a distinct physical quantity from the OAM of the modern theory. The approach of Ref.~\onlinecite{Orbitronics-PRL-2005-Shoucheng} is already of questionable validity for Si, where spin-orbit coupling is not negligible for hole systems, in fact it has been used very successfully for electrical spin manipulation in quantum computing. Such an approach would be entirely incorrect for Ge, GaAs, InAs and InSb, where spin-orbit coupling is inherently strong. Likewise, Ge has been used to achieve fast EDSR \cite{Hendrickx2020_NC} as well as two-qubit logic \cite{hendrickx2020fast} in hole qubits. This motivated us to cast a fresh glance at the OHE in spin-3/2 hole systems in the context of the modern theory.

An additional motivation is the fact that studies within the modern theory often focus on inversion breaking systems\cite{PhysRevB.107.094106,cullen2025quantum,OM-metal-PRB-2021-Xiaocong,Exp-graphene-OHE-arXiv-2022,Symmetry-PhysRevB.103.085113,TMDsoc-IOHE-PRB-Satpathy-2020,liu2025quantumOHE,PhysRevB.108.075427-Manchon}, which typically exhibit large values of the Berry curvature\cite{PhysRevB.108.075427-Manchon}. 
Yet, as is evident in this work, a large Berry curvature and OAM do not require inversion symmetry breaking \cite{PhysRevLett.130.116204}: they are large for spin-3/2 holes described by the Luttinger Hamiltonian, even when the spherical approximation is applied. The correct symmetry analysis for the Berry curvature was performed in Ref.~\onlinecite{PhysRevB.68.045327}. Additionally, for centrosymmetric, time-reversal-invariant systems in which the orbital Berry curvature is zero, such as those considered in Ref.~\onlinecite{han2023microscopic}, we find that while the conventional contribution to the OHE is zero the total OHE including the quantum correction can be nonzero.

Nevertheless it is important to keep in mind the limitations of effective mass studies, whose applicability is restricted to the conduction and valence band extrema. The orbital Hall conductivity for the full band structure of Ge and Si was recently calculated using ab initio techniques in Ref.~\onlinecite{IOHE-XIV-PRB-2021-Hyun-Woo}, while accounting fully for spin-orbit effects. { In the regions where numerical comparison is possible our results are approximately of the same order of magnitude, though it should be noted that the magnitude of our results grows much faster with the Fermi energy.} However, a direct comparison is challenging at the moment, since Ref.~\onlinecite{IOHE-XIV-PRB-2021-Hyun-Woo} used a basis of atomic OAM states, and the OAM operator entering the orbital current is the atomic OAM operator. In our evaluation, the OAM is computed using the modern theory and the effective mass envelope functions, incorporating additional quantum corrections which we recently shown to be vital in this formulation. 

These observations open up an important and interesting question for future research. At the moment two perspectives exist in the study of OAM in solids. One perspective, exemplified by Ref.~\onlinecite{IOHE-XIV-PRB-2021-Hyun-Woo}, takes as its starting point the OAM of atomic states and constructs the overall OAM response of a solid to an electric field. The other perspective is provided by the modern theory, whose Bloch function formulation is used in this work, and which can also be cast in terms of Wannier functions\cite{Theroy-OM-PRL-2007-Qian, Resta-PhysRevLett.95.137205, Vanderbilt_2018, Resta-PhysRevResearch.2.023139, Resta-PhysRevB.74.024408, thonhauser2011review, Rhonald-Rev}. The quantities calculated using these methodologies are clearly related and lead to the same observable. The modern theory and atomic OAM have produced consistent results for the orbital magnetisation for materials with their OAM concentrated about the atomic centre. \cite{OM-Berry-PRB-2016-Mokrousov} As such, since the modern theory accounts for all components of Bloch electron OAM not just the local circulation,\cite{Vanderbilt_2018, thonhauser2011review} the atomic OAM approach is assumed to be subsumed by the modern theory, with the exception of some 2D models in which in-plane OAM in the modern theory is forbidden. The exact relationship between the atomic OAM and modern theory has not been clarified to date. Determining where these perspectives intersect will be a task for future studies. Both approaches show that OAM conservation is not guaranteed \cite{CIAM-PRR-2020-Yuriy, Rhonald-Conservation-OMM, ODynamics-PRL-2022-Kyoung-Whan}, however, in the modern theory the conditions required for OAM conservation have been determined \cite{Rhonald-Conservation-OMM}. Since there is no orbital magneto-electric effect in the Si and Ge there will be no net torque on the OAM captured by the modern theory, and hence the OAM will be conserved in these materials. 
In this context we note that the fundamental definition of the magnetic moment has also come under scrutiny recently \cite{ado2025magnetic}. Although disorder can dominate the conventional contribution to the orbital Hall conductivity in certain systems\cite{veneri2024extrinsic,Hong-OHE-PRL}, we have ignored disorder effects in this work. In general for a full calculation of the orbital current the exact role of disorder is unknown at this point in time. As disorder is yet to be properly treated in a complete evaluation of the orbital current including quantum corrections beyond a simple relaxation time approximation \cite{liu2025quantumOHE}, we intend to address this in future work.

\subsection{Role of Zitterbewegung}
In this work we have focussed on intrinsic effects in the presence of an electric field. These stem from the inter-band part of the density matrix and are related to inter-band mixing induced by an electric field, in other words, Zitterbewegung. All contributions to the intrinsic orbital and spin currents arise due to interband elements of the nonequilibrium density matrix and Berry connection. In the absence of spin-orbit coupling the Luttinger parameters $\gamma_2$ and $\gamma_3$ vanish, as does the split-off energy gap, hence the split-off band will become degenerate with the light and heavy hole bands at $\bm k=0$. Thus, in the limit of vanishing spin-orbit coupling, all states are degenerate so the interband elements of the non-equilibrium density matrix vanish. Additionally, the Hamiltonian becomes proportional to the identity matrix and the Berry connection also vanishes. In this limit the effects discussed in this work will vanish (although in practice the OHE and SHE may have small contributions due to Zitterbewegung involving the conduction band, which is not included here, as explained below). Nevertheless, in general spin-orbit coupling is not necessary to generate an OHE. The key mechanism is Zitterbewegung \cite{cullen2025quantum, liu2025quantumOHE, atencia2025quantum}, which can be associated with spin-orbit coupling as it is here, with pseudospin dynamics as in graphene, or with some other phenomenon in more complex band structures.

\subsection{Limits of applicability of the $\bm k \cdot \bm p$ method} 
The $\bm k \cdot \bm p$ models used in this work are low $\bm k$ expansions about the valence band centre \cite{Roland}. The Berry connection and curvature play crucial roles in the modern theory of orbital magnetisation. These quantities often require more sophisticated models to effectively capture the behaviour of the Bloch wavefunction. However, at the low carrier densities we are interested in to describe semiconductor transport, the most important quantity that needs to be captured is the Berry curvature monopole at the band centre which decays as $1/k^2$. The $\bm k\cdot \bm p$ Hamiltonians we use here are expanded in terms of the Bloch wavefunction at the band edge ($\bm k = \bm k_0$)\cite{chow&koch,marder2010condensed}. This expansion takes the form $|u_{m\bm k}\rangle=\sum_n c_{nm \bm k}|u_{n\bm k_0}\rangle$, and the derivative is expressed as $\partial/\partial k^\alpha|u_{m\bm k}\rangle=\sum_n (\partial/\partial k^\alpha c_{nm \bm k})|u_{n\bm k_0}\rangle$ which enters the Berry connection and the Berry curvature. These wavefunctions are then approximated by retaining only a few bands that are close in energy. This is an excellent approximation for the small Fermi energies considered in this work $E_{\text{F}}<10$ meV. Since these Fermi energies also correspond to the most experimentally accessible carrier densities there is no loss of generality. Additionally, effectively implementing the modern theory of orbital magnetism through more sophisticated ab-initio models can be very expensive computationally, often requiring a large number of atomic orbitals spanning a substantial energy range. \cite{vidarte2025real} 

Including coupling to further bands will give corrections to the values calculated in this work, and these corrections will depend on the ratio of the $\bm k \cdot \bm p$ coupling strength to the direct energy gap between the bands. The difference in the orbital Hall conductivity between the $4\times4$ and $6\times6$ models is $12\%$ for Ge, as is shown in Fig.~\ref{Fig:OHE_Ge_compare}. An important point to note is that the conduction band coupling elements may be fairly sizable and could introduce non-negligible corrections. Values for the direct conduction-valence band gap $\delta_{\text{c}}$ are given in Tab.~\ref{tab:split-off}. For the carrier densities considered in this work the conduction band coupling to band gap ratio can be up to: $18\%$ in Si, $23\%$ in GaAs, $61\%$ in Ge, $71\%$ in InAs and $108\%$ in InSb. The coupling strength to gap ratios for all other bands are much smaller and should introduce negligible corrections \cite{Roland}. The conduction band coupling to gap ratio compared to the split-off band coupling ratio will be: $\sim 3$ times smaller in Si, similar in GaAs, and $\sim 2 - 5$ times greater in Ge, InAs, and InSb. Based on these figures, we estimate that a calculation using the $8\times8$ Luttinger model including the conduction band for Si and GaAs is likely to introduce a correction to the orbital Hall conductivity of $\sim 10\%$, in Ge and InAs we expect the correction to be larger $\sim30-40\%$. For InSb there could be a correction as large as $\sim80\%$ from including the conduction band, given the large values calculated for InSb such a correction would still yield a sizable orbital Hall conductivity. With this in mind the results for InSb should be viewed as indicative, with further analysis required. Nevertheless, very generally, the correction obtained by including the split-off band is only a fraction of the orbital Hall conductivity arising from the original $4\times4$ model, and therefore even when considering the limitations of our approach, the core conclusions of this work are still valid: a large orbital Hall effect exists in $p$-type semiconductors, and it is dominated by the quantum correction.

\section{Conclusions}

We have demonstrated that hole-doped semiconductors Ge, Si, GaAs, InAs and InSb exhibit a large orbital Hall effect and are suitable platforms for orbitronic applications. The OHE response stems from Zitterbewegung induced by spin-orbit coupling, is dominated by the quantum correction, and exceeds the spin-Hall effect by 2-3 orders of magnitude. The absence of spin and orbital Edelstein effects suggests both Ge and Si as platforms for testing the strength of the orbital Hall torque. Finally, our work shows that the OAM and OHE calculated within the framework of the modern theory of orbital magnetisation are strong for the Luttinger Hamiltonian even in the spherical approximation, providing an example of strong orbital dynamics in an inversion-symmetric system.

\section{Data availability}
The authors declare that the data supporting the findings of this study are available within the paper and its supplementary information file.

\section{Author contributions}
J.~H.~C and Z.~W performed the orbital and spin Hall effect calculations. J.~H.~C made the figures. J.~H.~C and D.~C wrote the manuscript. D.~C Supervised the project.

\section{Competeing interests}
The authors declare no competing interests.

\section{Acknowledgements}. This work is supported by the Australian Research Council Discovery Project DP2401062 and Future Fellowship FT190100062. We are very grateful to Tatiana Rappoport, Hyun Woo Lee, Kyoung-Whan Kim, Jung Hoon Han, Francesco Borsoi, Joe Salfi, and Henri Jaffres for stimulating discussions.


%

\section{Figure captions}
Figure 1: Dispersion for holes in Ge, showing the heavy (HH), light (LH) and split-off (SO) hole bands. In this figure the bands have been inverted so the energy is positive. Here we have chosen a Fermi energy of $10$ meV, which is the largest Fermi energy we consider in this work.

Figure 2: The orbital Hall conductivity vs the Fermi energy in Ge in the spherical approximation. We have also plotted the conventional part of the orbital Hall conductivity $\sigma_{\text{conv}}$ along with the quantum corrections $\Delta\sigma$. Here we use the Luttinger parameters $\gamma_1=13.38$ and $\bar{\gamma}=4.97$.

Figure 3: The proper spin Hall conductivity vs the Fermi energy in for four semiconductors in the spherical approximation. Here we use the Luttinger parameters from Ref.~\onlinecite{Roland}. Note the difference of three orders of magnitude in the scale of the $y$-axis as compared to the orbital Hall conductivity.

Figure 4: The orbital Hall conductivity vs the Fermi energy in Ge for a) the $4\times4$ Luttinger Hamiltonian and b) the $6\times6$ Luttinger Hamiltonian without the spherical approximation. We have plotted the conventional part of the orbital Hall conductivity $\sigma_{\text{conv}}$ along with the quantum corrections $\Delta\sigma$. Here we use $\gamma_1=13.38$, $\gamma_2=4.24$ and $\gamma_3=5.69$.

Figure 5: The orbital Hall conductivity vs Fermi energy for various hole-doped semiconductors calculated using the $6\times6$ Luttinger Hamiltonian. Luttinger parameters have been taken from Ref.~\onlinecite{Roland}.

Figure 6: Diagram of the orbital Hall torque in a Ge/Co heterostructure. Here the applied electric field generates transverse orbital currents via the intrinsic orbital Hall effect. The orbital current will generate an orbital accumulation. The orbital angular momentum is then converted into spin which generates a torque on the magnetisation in the Co layer. A barrier layer may be used to assist with orbital-to-spin conversion.

\end{document}